\documentclass[preprint2]{aastex}

\shorttitle{Optical tomography of V926 Sco}
\shortauthors{Connolly et al.}
\def\etal{{\em et al.}\ }

\begin{document}


\title{Variability of the accretion disk of V926 Sco inferred
from tomographic analysis.}


\author{S.D. Connolly}
\affil{ University of Southampton, Highfield, Southampton, S017 1BJ, UK} 
              \email{sdc1g08@soton.ac.u}
\author{C.S. Peris\altaffilmark{1}}
\affil{ Department of Physics, Northeastern University, Boston, MA 02115, USA}
        \email{peris.c@husky.neu.edu}
\author{S.D.Vrtilek}
\affil{Harvard-Smithsonian Center for Astrophysics, Cambridge, MA 02138, USA}
	\email{svrtilek@cfa.harvard.edu}


\altaffiltext{1}{Harvard-Smithsonian Center for Astrophysics, Cambridge, MA 02138, USA.
cperis@cfa.harvard.edu}


\begin{abstract}
We present phase-resolved spectroscopic observations of the low-mass
X-ray binary V926 Sco (4U 1735-44), 
covering the orbital period of 0.23d, obtained with the Walter Baade 6.5m Magellan 
Telescope at the Las Campanas Observatory in June 2010 and June 2011.  
We use H$\alpha$ radial velocities to
derive a systemic velocity of -109$\pm$4km s$^{-1}$.
The FWHM of the lines observed in common with previous authors
are significantly lower during our observations suggesting much reduced velocities in the
system. The equivalent width of the Bowen fluorescence
lines with respect to He~II $\lambda$4686 are factors of two or
more lower
during our observations in comparison to those previously 
reported for the system,
suggesting reduced irradiation of the secondary.
Doppler and modulation tomography of H$\alpha$ and He~II $\lambda$4686  show assymmetric emission that can
be attributed to a bulge in the accretion disk, as inferred from He~II
 observations
by previous authors.
The X-ray fluxes from the source at times concurrent with
the optical observations 
are significantly lower during our observations than during
optical observations taken in 2003. 
 We suggest that the system is in a lower accretion state compared to
earlier observations; this explains  
both the lower velocities observed from the disk and the reduction
of emission due to Bowen fluorescence detected from the
secondary.
\end{abstract}


 \keywords{accretion, ccretion disks --
                stars: individual (V926Sco)
		stars: neutron
               }



\section{Introduction}

V926 Sco (4U 1735-444) is a low mass X-ray binary (LMXB) that is persistent in X-rays.  
The shape of its X-ray color-color diagram caused Hasinger \& van der Klis (1989) to classify it as an atoll source.
The system has an orbital period of 4.65 hours, discovered through optical photometry 
(Corbet~\etal~1986, Pederson~\etal~1981), which showed a shallow 
sinusoidal variation in the light curve, interpreted as due to the varying aspect 
of the X-ray heated secondary object (e.g. van Paradijs~\etal~1988).   
Although sinusoidal variation can also be produced by asymmetries in the disk,
for example through the varying visibility of an irradiated inner disk bulge
(Hellier \& Mason 1989) or by superhumps (Haswell~\etal~2001),
the variation was attributed to the donor by Casares~\etal~(2006).

A periodic variation in the H$\alpha$ line observed by Smale~\etal~(1984) was 
confirmed by
Smale \& Corbet~(1991) and attributed
to varying emission originating from a bulge or splash region at or near the point on 
the disk at which the gas stream impacts the outer rim.  Augusteijn~\etal~(1998) 
found similar variation in He II $\lambda$4686 
and the blend of N~III and C~III emission lines ($\lambda\lambda$4634-4651) produced by 
Bowen fluorescence, and attributed both to a disk bulge.

Doppler tomography (Horne \& Marsh~1986) was carried out on V926 Sco by
Casares~\etal~(2006), using data taken in June 2003 with the FORS2 Spectrograph 
on the 8.2m Yepun Telescope at the Observatorio Monte Paranal.
Tomography of He II $\lambda$4686 showed an extended area of bright emission on one side of the disk,
suggesting that the variation in this emission does arise from a bulge in the accretion disk,
consistent with Smale \& Corbet~(1991).
However, Casares~\etal~(2006) found that N~III $\lambda$4640
 emission was coincident with the estimated velocity of the donor star, as opposed to a disk bulge. 
They attributed this to fluorescence of the donor, irradiated by UV photons from the hot inner disk, 
as originally suggested for Sco X-1, along
with several other X-ray binaries, by McClintock, Canizares \& Tartar~(1975).

\begin{figure*}
 \includegraphics[width=170mm, height=110mm]{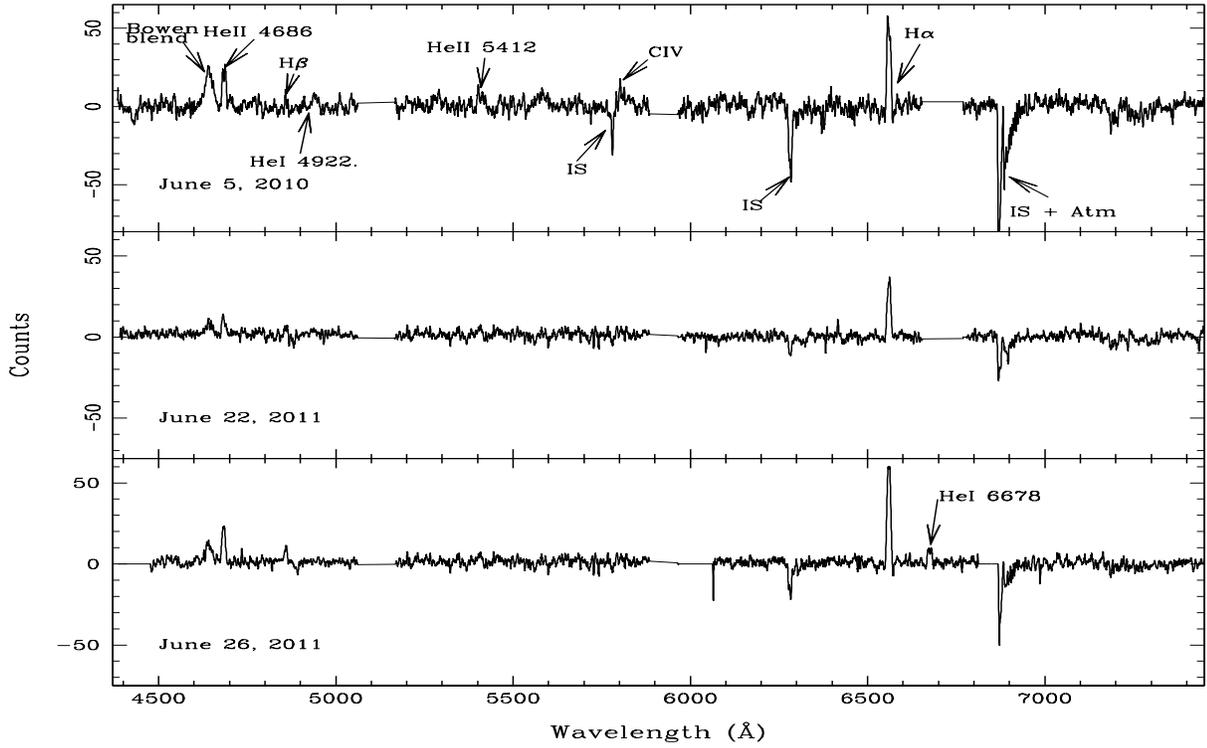}
 \caption{The smoothed average continuum-subtracted spectra of V926 Sco for the nights of June 5, 2010, June 22, 2011, and June 26, 2011 show H$\alpha$, H$\beta$,  He~II $\lambda$4686 \& $\lambda$5412, C~IV $\lambda$5807, He~I $\lambda$4922 \& $\lambda$6678, and the blend of N~III and 
C~III emission ($\lambda\lambda$4634-4651) comprising the Bowen complex. Interstellar lines are marked as IS.}
\label{avspec}
\end{figure*}

Our spectra cover nearly twice the bandwidth of Casares~\etal~(2006) including in particular the H$\alpha$ line for
which no previous tomographic study has been done and which
is particularly useful for studying emission from the disk.  In addition to
Doppler tomography we also undertake {\it modulation} tomography 
(Steeghs~\etal~2003)
which allows us to take into consideration variations in the brightness of the components of the system which are harmonic with the orbital period. As Doppler tomography is limited by the assumption that the brightness of the components of a system do not vary, its combination with modulation tomography allows more accurate interpretation of the velocity maps of a system.

In section 2 we describe the optical observations obtained and used for this study.  In section 3 we present our analysis
of the spectral features.  In section 4 we present Doppler and modulation tomography of the lines and end with a summary
and conclusions in section 5.

\section{Observations}


Observations of V926 Sco were carried out on the nights of 2010 June 5,7 and 2011 June 22, 26, using 
the {\it IMACS} spectrograph on the Walter Baade 6.5 m Magellan telescope at the Las Campanas 
Observatory. A long-slit diffraction grating with 600 lines/mm at a tilt angle of  $11.23^o$ was 
used, giving a wavelength range of approximately 4449-7581${\AA}$
(excluding small gaps due to CCD chip edges), with $37$ km s$^{-1}$ (FWHM) resolution.
A slit width of 0.9 arc seconds was used on 2011 June 22 and of 0.7 arc seconds on 
2011 June 26 and 2010 June 5,7. 60 useful spectra were obtained, with exposures 
of 420-600 seconds, covering a total of approximately 1.5 orbital periods
(Table 1). HeNeAr comparison lamp arcs were taken after approximately every three exposures.
The images were corrected for bias and flat-fielded, then the spectra were extracted using
{\it IRAF} optimal extraction for a weak spectrum, as described by Massey, Valdes,
\& Barnes~(1992), to give the best possible signal to noise ratio.
Anomalies due to cosmic rays and CCD errors were removed. The spectra were
 wavelength-calibrated using the time-nearest HeNeAr comparison arcs, with a
separate
second order polynomial fit of each of the four CCD chips across which the spectra
 were spread. In each case, the RMS scatter was $<0.01 {\AA}$.
The flux standard Feige 110 was observed on the night of June 22, 2011.  However,
since none of our nights were photometric, 
we plot only relative intensities and give EWs in instrument counts.

\begin{table*}
      \caption{Observation Log.}
         \label{tab1}
         \begin{tabular}{lccccccc}
            \noalign{\smallskip}
         &    &     &          &        &Phase        & No. of &  \\
 Date        & UTC (start)    & UTC (end)    &  MJD (start)        & MJD (end)        &coverage        & spectra & Exposure(s) \\
            \hline
            2010 June 5          & 00:39:20    & 04:18:51    & 55353.0273    & 55353.1798    & 0.22-1.01        & 15         & 420-600    \\
            2010 June 7          & 01:00:45    & 01:42:32    & 55355.0422  & 55355.0712    & 0.99-1.14        &  4          & 600            \\
            2011 June 22        & 05:54:58    & 09:02:57    & 55735.2465    & 55735.3770    & 0.76-0.44        & 15         & 600            \\
            2011 June 26        & 01:59:24    & 07:08:33    & 55739.0829    & 55739.2976    & 0.18-1.29        & 26         & 600            \\
            \hline
         \end{tabular}
   \end{table*}

 \begin{table}
      \caption{Emission Line Parameters}
         \label{tab2}
         \begin{tabular}{lccc}
            \noalign{\smallskip}
            \hline
            Line      &  FWHM & EW&Centroid \\
                 & (km s$^{-1}$)&(Inst Cts)&($\AA$)\\
            \hline
06/05/2010&&&\\
                Bowen blend &956$\pm$31&23.1$\pm$1.5&4641.1\\
                He~II $\lambda$4686 &420$\pm$20&23.5$\pm$2.0&4685.6\\
                H$\beta$&---&---&4860.9\\
                H$\alpha$&417$\pm$10&60.6$\pm$2.6&6561.2\\
                He~I $\lambda$6678&---&---&6678.0\\
\hline
06/07/2010&&&\\
                Bowen blend &710$\pm$39&23.0$\pm$2.4&4640.1\\
                He~II $\lambda$4686 &425$\pm$23&41.3$\pm$4.3&4686.1\\
                H$\beta$&---&---&4860.9\\
                H$\alpha$&419$\pm$10&69.2$\pm$5.1&6562.3\\
                He~I $\lambda$6678&---&---&6678.0\\
\hline
06/22/2011&&&\\
                Bowen blend &770$\pm$63&6.9$\pm$1.0&4641.9\\
                He~II $\lambda$4686 &380$\pm$25&11.5$\pm$1.4&4684.6\\
                H$\beta$&313$\pm$65&4.2$\pm$1.6&4860.3\\
                H$\alpha$&394$\pm$10&37.3$\pm$1.8&6561.8\\
                He~I $\lambda$6678&---&---&6679.9\\
\hline
06/26/2011&&&\\
                Bowen blend &928$\pm$33&11.3$\pm$1.0&4641.3\\
                He~II $\lambda$4686 &383$\pm$11&24.3$\pm$1.3&4684.7\\
                H$\beta$&283$\pm$24&8.9$\pm$1.5&4859.8\\
                H$\alpha$&379$\pm$9&65.1$\pm$1.6&6561.3\\
                He~I $\lambda$6678&483$\pm$28&10.7$\pm$1.2&6676.8\\
            \noalign{\smallskip}
            \hline
         \end{tabular}
   \end{table}

\section{Spectral features and profiles}

The spectra of V926 Sco averaged over the nights of June 5, 2010 and
June 22 and June 26 2011 are presented in Fig. \ref{avspec}. 
Each contains relatively strong H$\alpha$ emission in addition to 
weaker lines of H$\beta$, He~II $\lambda$4686 \& 
$\lambda$5412, HeI $\lambda$4922, and a blend of N III and C III emission lines 
($\lambda\lambda$4641-4651) attributed to Bowen fluorescence,
 C~IV $\lambda$5807,
and He~I $\lambda$6678. These emission features are consistent with those 
found by previous spectroscopic studies (Casares~\etal~2006), 
Cowley~\etal 2003, Augusteijn~\etal~1998,
Smale~\& Corbet~1991).
Several absorption features attributed to interstellar and 
atmospheric absorption
are also present.

\begin{figure*}
\includegraphics[width=80mm]{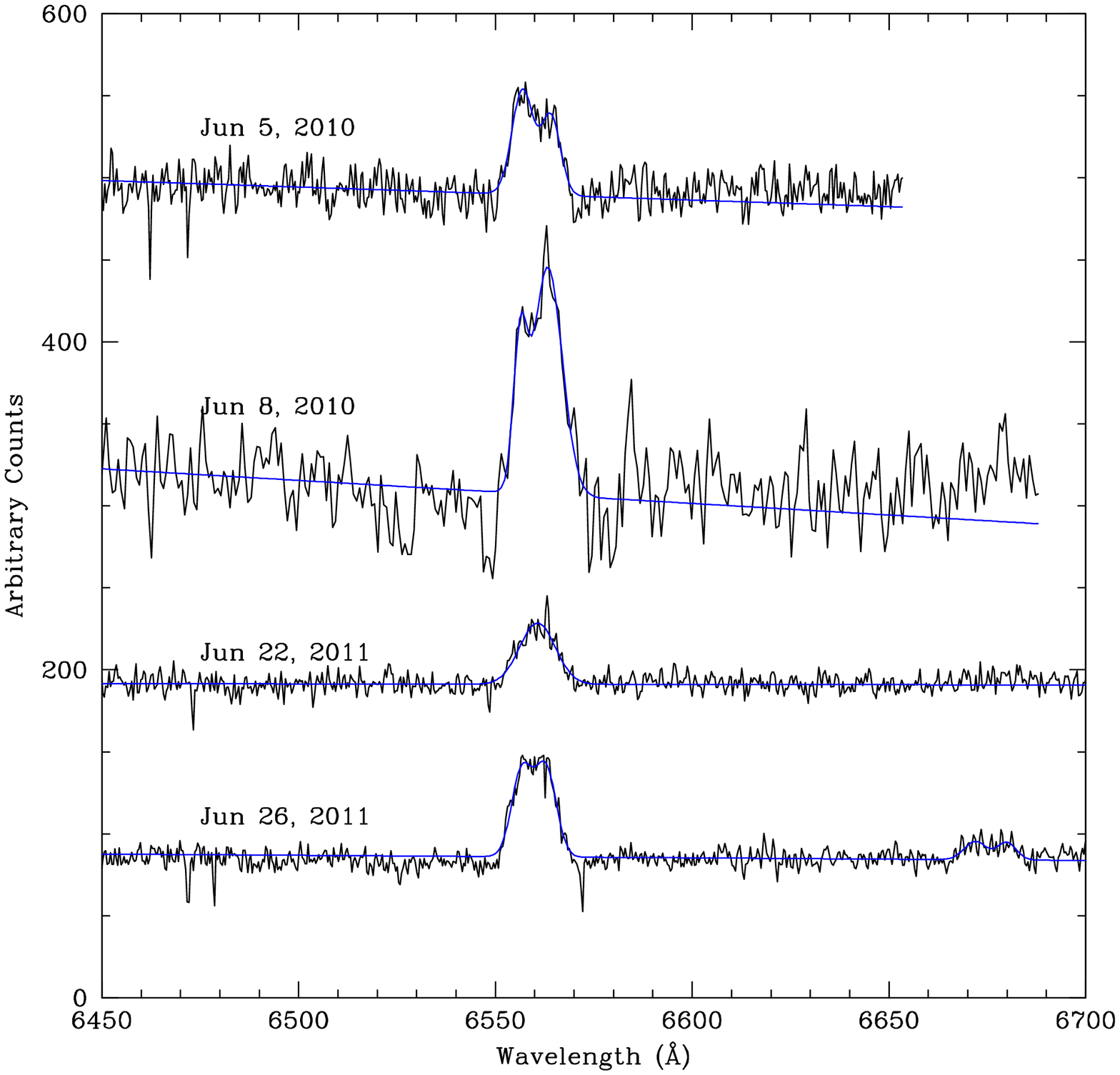}
 \includegraphics[width=80mm]{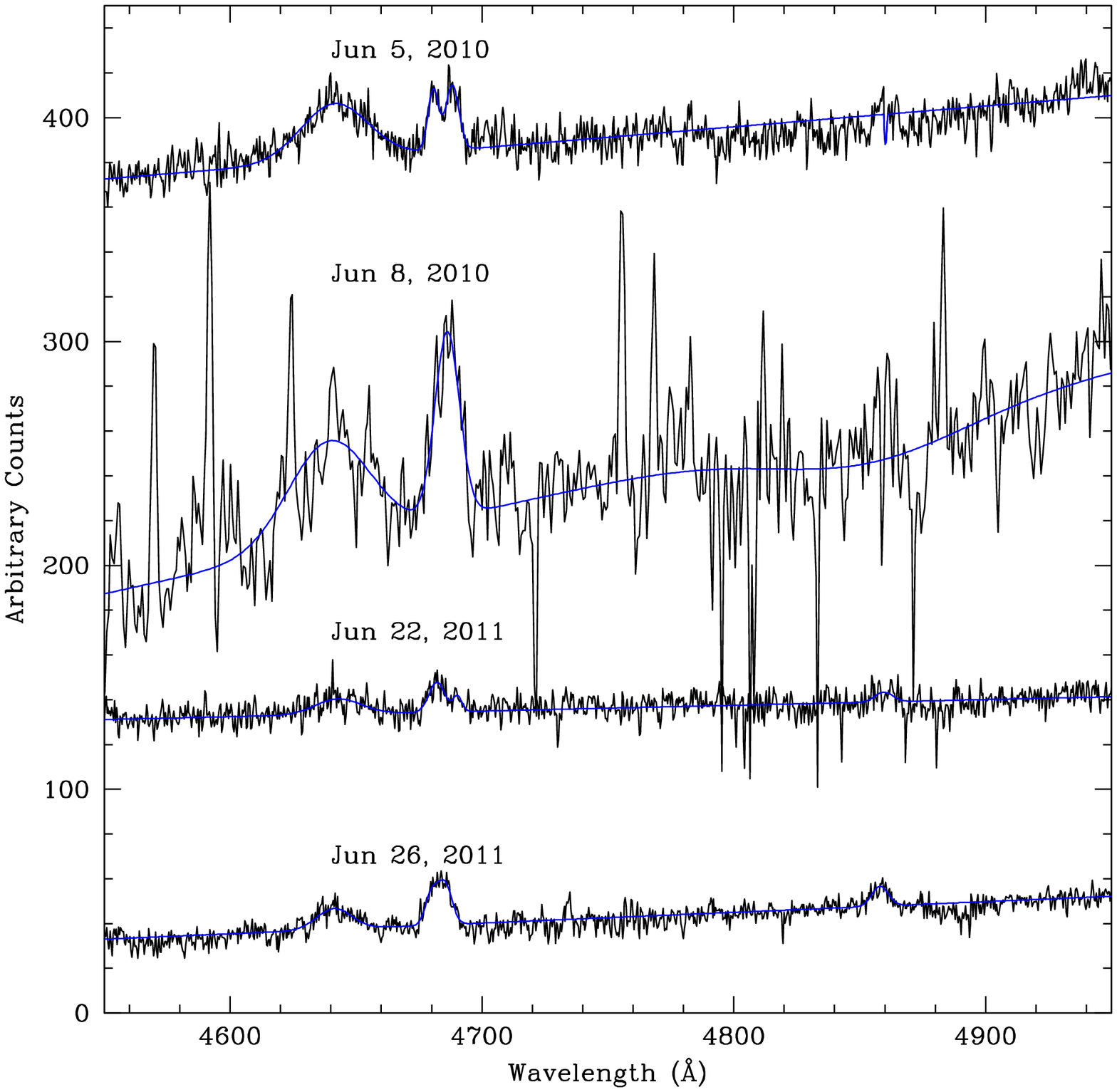}
 \hspace{0.7cm}
 \caption{Gaussian fits to H$\alpha$, He~I $\lambda$6678,
the Bowen complex, He~II $\lambda$4686, and H$\beta$ for each of the
nights
listed in Table 1. Values of the fitted parameters are listed in Table 2.
{\bf Left (a):} H$\alpha$ and He~I $\lambda$6678 both showed the classic
double peak expected from emission from an accretion disk and required two
gaussians for good fits.
{\bf Right (b):}
Fits using two gaussians for the Bowen blend: $\lambda$4634 and
$\lambda$4641. He~II $\lambda$4686 and H$\beta$ also displayed
double peaks and required two
gaussians.}
\label{fit1}
\end{figure*}

\begin{figure}
 \includegraphics[width=75mm, height=110mm]{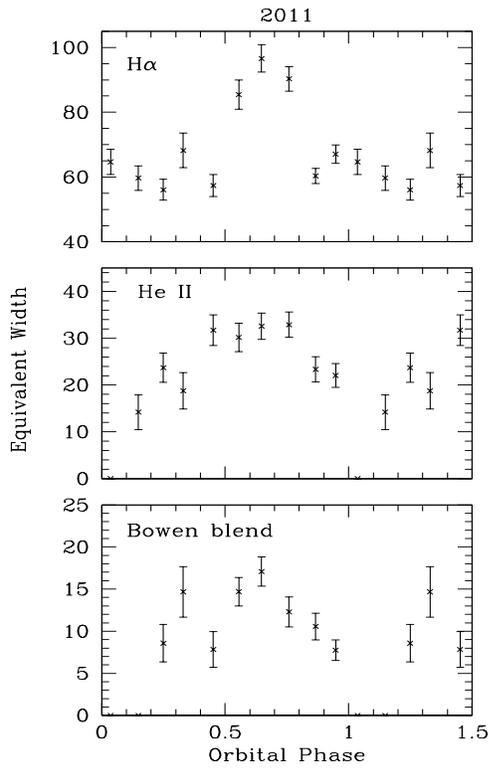}
\caption{Equivalent widths for the 2011 data in 10 phase bins. The phase
of peak equivalent width (0.7)
is consistent with emission from the
disk bulge. The 2010 data do not have sufficient
phase coverage to warrant phase resolved values.}
\label{ew}
\end{figure}

Figs. \ref{phasecur} show the evolution of the main spectral features over
the orbital period.  We use the spectroscopic ephemeris of Casares~\etal~(2006)
to determine orbital phases:

T$_o$(HJD) = 2452813.495(3) + 0.19383351(32)E

\noindent
where T$_o$ is defined as the compact object superior conjunction. 
H$\alpha$ and He~II $\lambda$4686
display
the classic double peak variation, with the peak shifting from
red to blue over the orbital period as expected for emission
from an accretion disk.
The Bowen complex varies in strength over the orbit but does
not show the double peak behavior.
In order to determine the relative contribution of the spectral
features, we fit the average nightly profiles with Gaussians.  For the
Bowen blend we used Gaussians representing the N~III transitions
($\lambda\lambda$4634,4641,4642)
and C~III transitions
($\lambda\lambda$4647,4651,4652). 
For H$\alpha$, He~II $\lambda$4686, and He~I $\lambda$6678 we
used two Gaussians each to
represent the blue and red shifted emission from the disk.
 For all lines we used single Gaussians (for Bowen this was at $\lambda$4641) 
to determine
the instrumental EW.
The fits are plotted in Fig. \ref{fit1} and values for the fitted parameters are listed in Table 2. 
For the 2011 data, where we had full phase coverage, we also
calculated equivalent widths as a function of orbital phase
(Fig. \ref{ew}).  The equivalent widths increase between phases
0.5-0.8 consistent with emission from the disk bulge. 
The source appears to be reduced in intensity from 2010
to 2011, however since we do not have absolute fluxes we
cannot determine this.  We do find that the equivalent width
of the Bowen complex is significantly reduced from 2010 to 2011.
Since the 2010 phase coverage is poor, for the Bowen complex we
construct tomograms only for the 2011 data.
The reduction in flux observed for June 22, 2011 compared to June 26, 2011 is
attributed to the larger slit size used due to poor seeing.
The increase in slit size also reduced our spectral resolution;
hence for our tomographic analysis of H$\alpha$ and He~II($\lambda$4686)
we use data from June 5, 2010 and June 26, 2011.
The four spectra obtained on June 7, 2010 were used only to
complete the radial velocity curve.

\begin{figure}
\includegraphics[width=80mm]{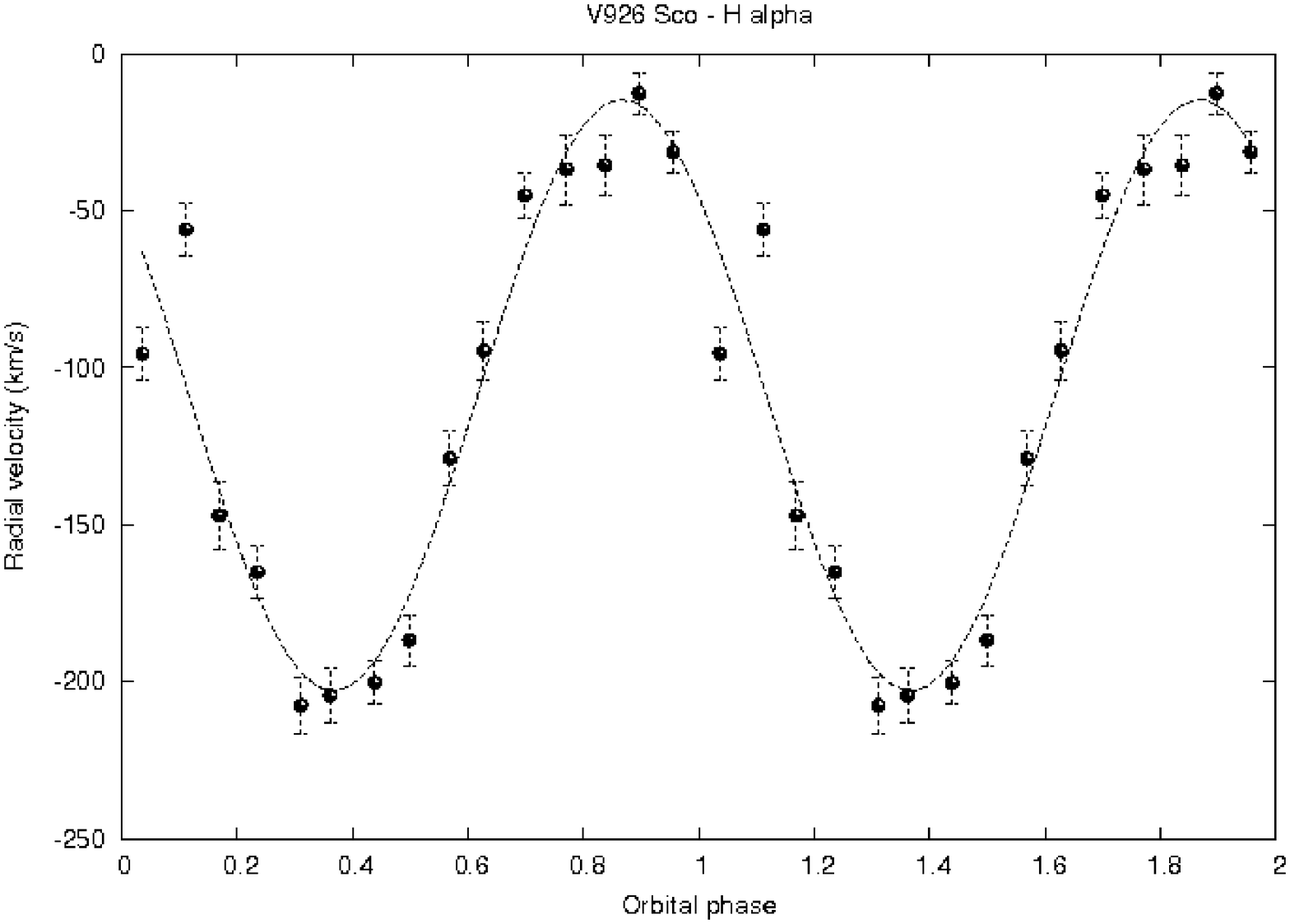}
\includegraphics[width=80mm]{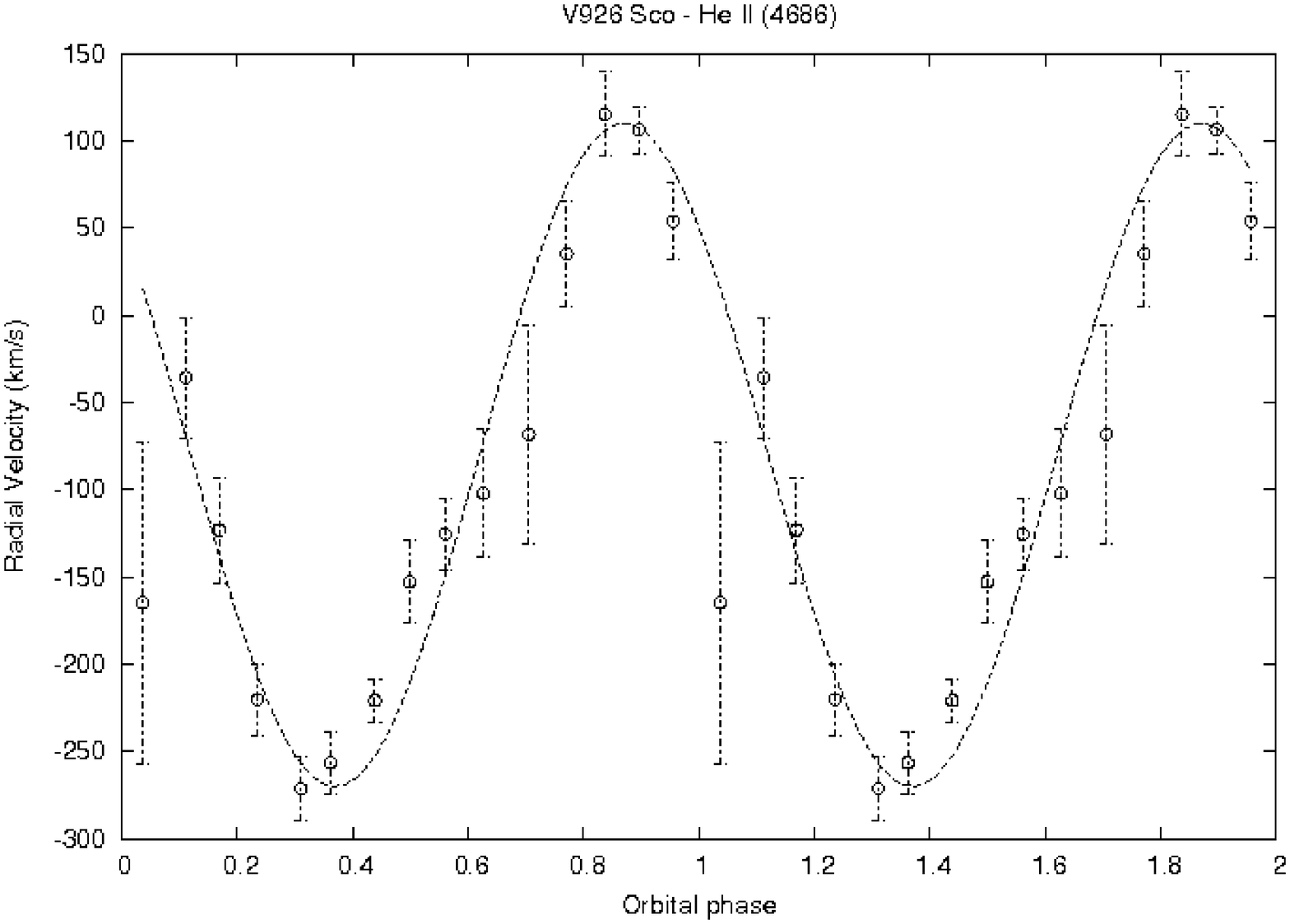}
 \hspace{0.7cm}
\caption{Radial velocity curves for H$\alpha$ and He~II($\lambda$4686) using the spectroscopic
ephemerides of Casares~\etal~(2006). 1$\sigma$ errors and the best-fit sinusoid are indicated.
The derived systemic velocity using H$\alpha$ is 109$\pm$4 km s$^{-1}$.}
\label{radvel}
\end{figure}

Figure \ref{radvel} shows the radial velocities for H$\alpha$
and He~II($\lambda$4686) 
obtained by cross-correlating all 60 individual spectra with Gaussians of FWHM
as listed in Table 2.
We used the spectrosopic ephemerides of Casares~\etal~(2006) 
as listed above.  The curves are in antiphase with 
that expected from the secondary and consistent with emission
from the disk.  Our best-fit
sine-wave to the H$\alpha$ data give us a systemic velocity of -109$\pm$4km s$^{-1}$ with a
semi-amplitude of 95$\pm5$km s$^{-1}$.
Our systemic velocity is consistent (within 2$\sigma$) with the value 
(121$\pm$7km s$^{-1}$) found by Casares~\etal~(2006) when using the wings of 
their HeII $\lambda$4686 profile (which are expected to follow the motion of the compact star) and our errors are a factor of two smaller.  Our HeII $\lambda$4686 is 
weak compared to H$\alpha$ but its radial velocity curve peaks at the same phase as H$\alpha$.

\section{Tomography}

Tomography is an imaging technique allowing two-dimensional velocity-space maps of a system 
to be reconstructed from spectra taken at multiple orbital phases. Spectra are assumed to 
be one-dimensional projections of the system in velocity space at a given phase. Under this 
assumption, if an accurate ephemeris is known an inversion technique can be used to produce 
possible fits to the data. In this case, a reduced $\chi^2$ test was used to modify an 
arbitrary starting image (e.g. a uniform or gaussian distribution) such that the predicted 
data from this image fit the real data. Due to the large number of possible fits for a given 
value of $\chi^2$ , the {\it `Maximum Entropy Method'} (MEM) is employed to select the image 
which is most likely to be accurate; the image with the highest entropy is chosen at each 
iteration, on the assumption that a higher entropy corresponds to a smoother and therefore 
more physically realistic image of the system (Narayan \& Nityananda 1986). For 
a complete description of imaging accretion disks using Doppler tomography, 
see Marsh \& Horne (1988).

Modulation tomography produces additional velocity maps showing the magnitude of periodic, sinusoidal modulations in the brightness of the structural features of the system, in addition to the velocity maps of the time-averaged brightness seen in Doppler tomography. This allows variations in the brightness of a system to be taken into account when interpreting the time-averaged velocity maps. Although modulation tomography allows more accurate interpretation of velocity data than Doppler mapping alone, which assumes emission to be constant over the orbital period, data with a higher signal to noise ratio is required.
For a more detailed description of modulation tomography of emission lines, 
see Steeghs (2003).

\begin{figure*}
\includegraphics[width=53mm]{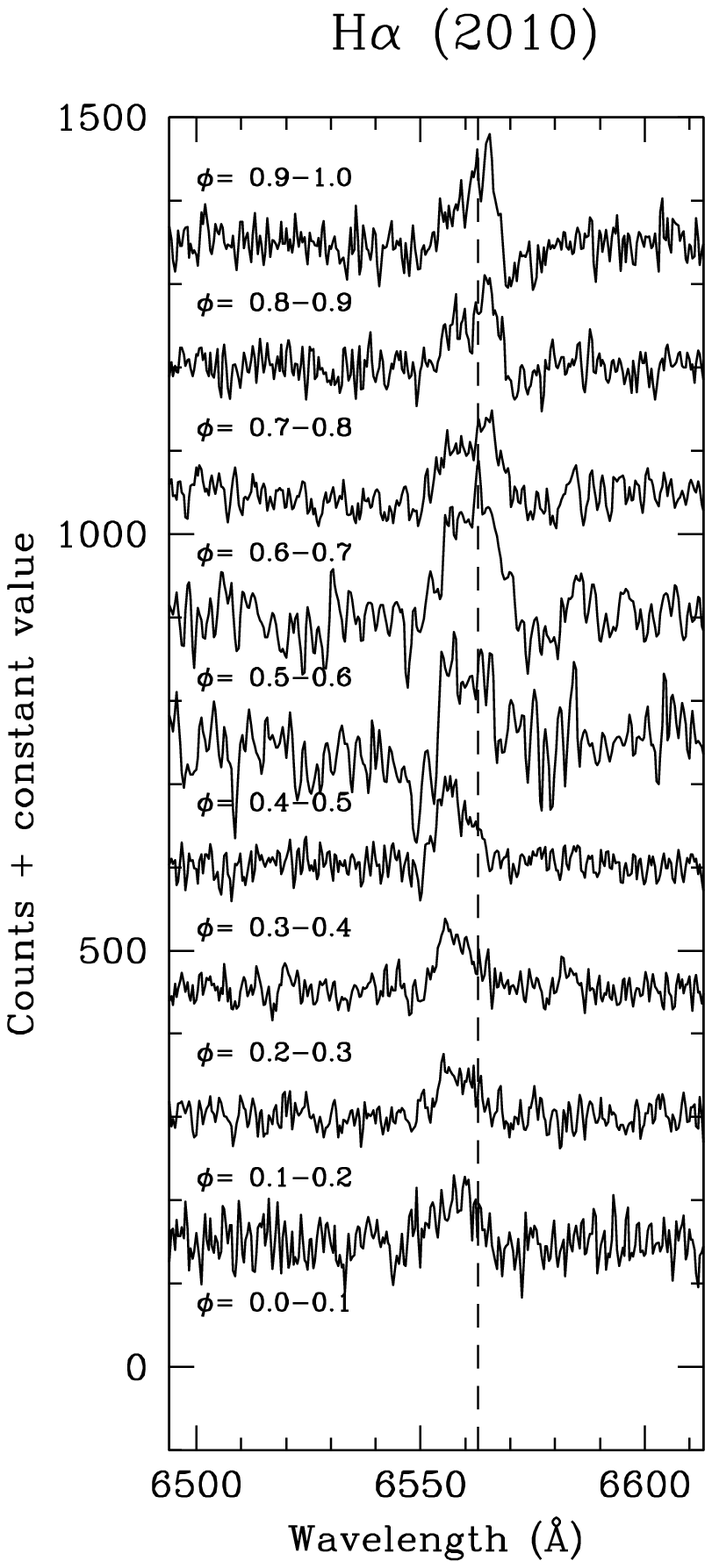}
\includegraphics[width=53mm]{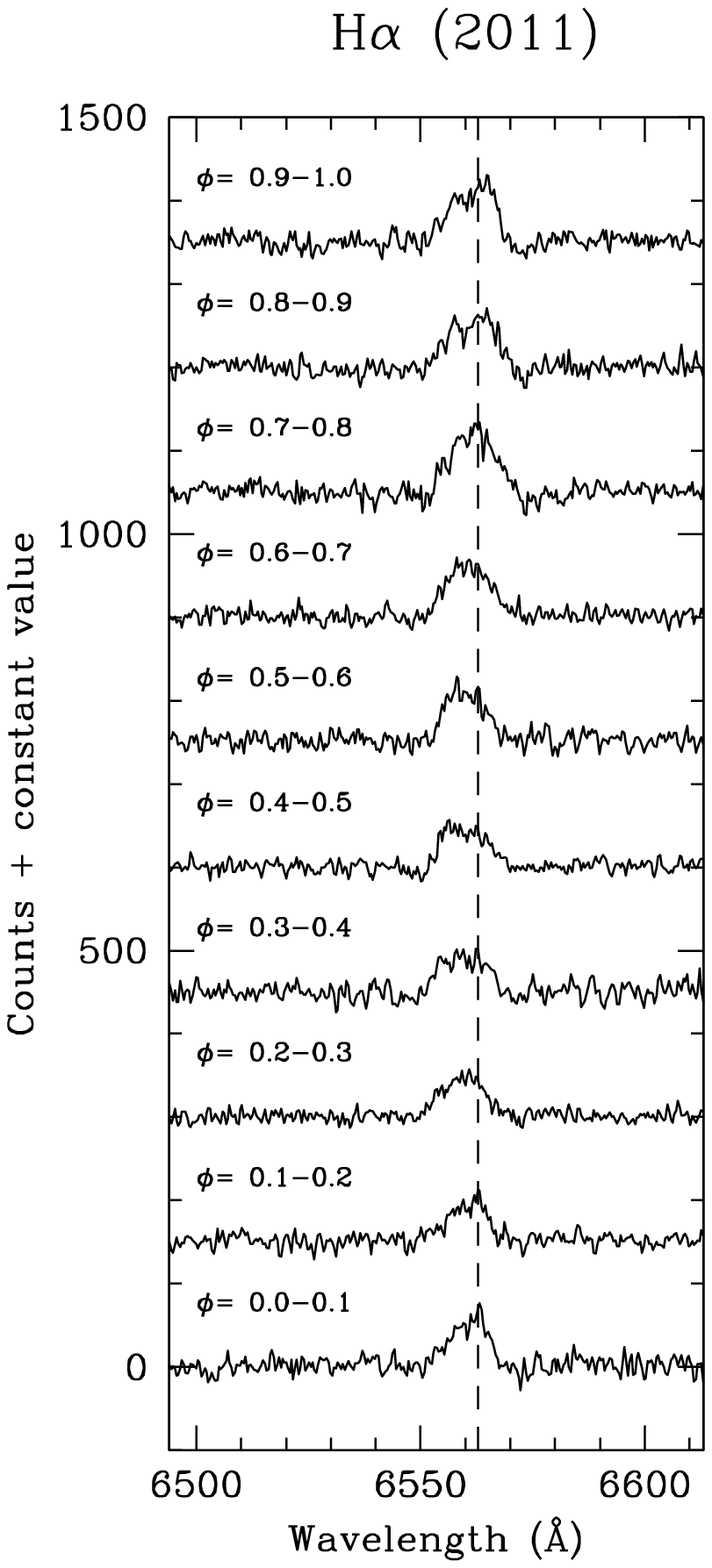}
\includegraphics[width=53mm]{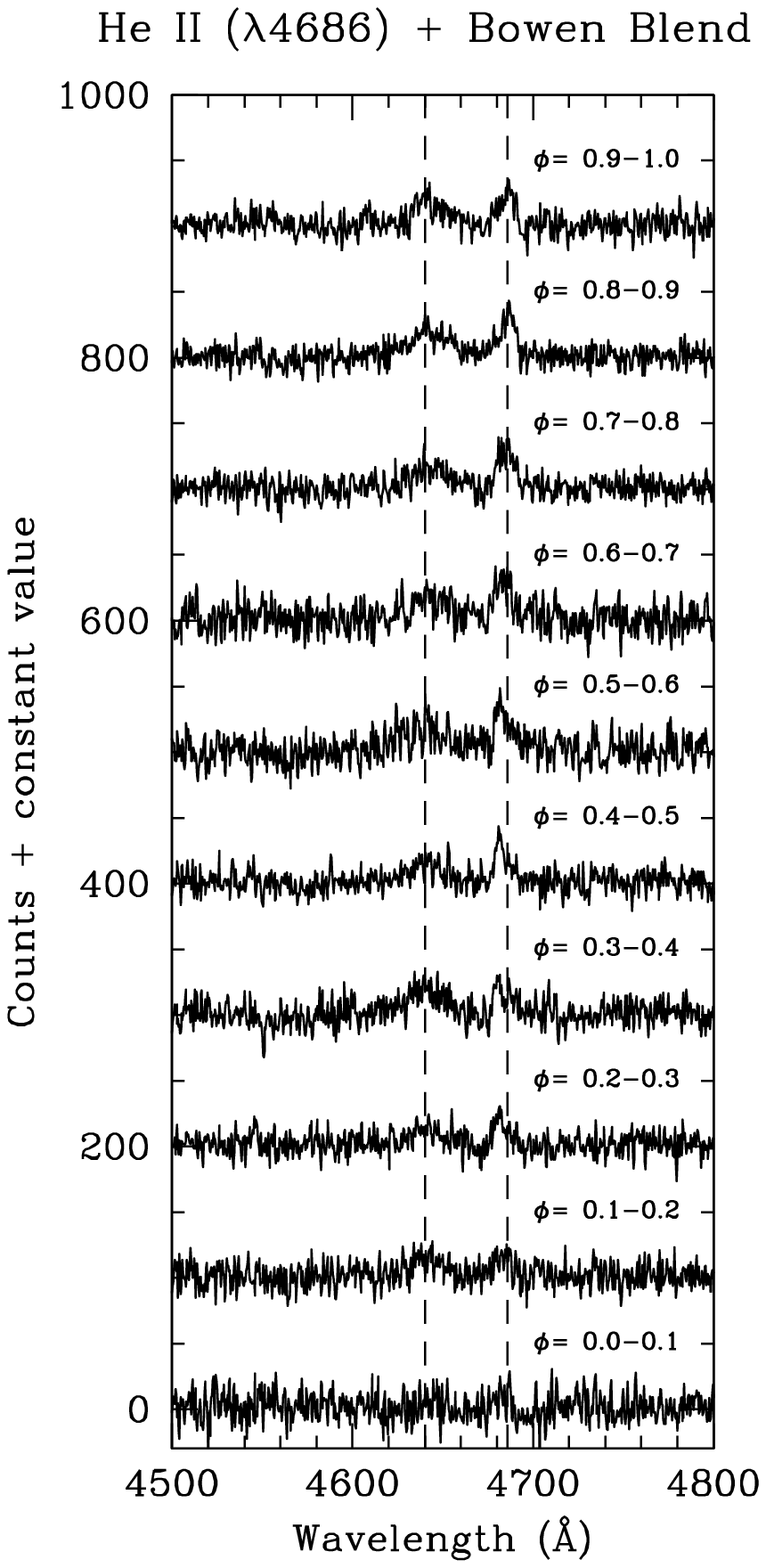}
\hspace{0.7cm}
\caption{{\bf Left (a):} H$\alpha$ spectra from 2010 in 10 phase bins, showing the
evolution of the lines over the orbital period. The dashed line marks
the laboratory wavelength of $\lambda$6562.8.
{\bf Center (b):} As in (a) for June 26,2011.
{\bf Left (c):} 2011 data of and He II and the Bowen blend in 10 phase bins, showing the
evolution of the lines over the orbital period. The dashed lines represent
$\lambda$4640 and $\lambda$4686.}
\label{phasecur}
\end{figure*}

\begin{figure*}
\includegraphics[width=80mm]{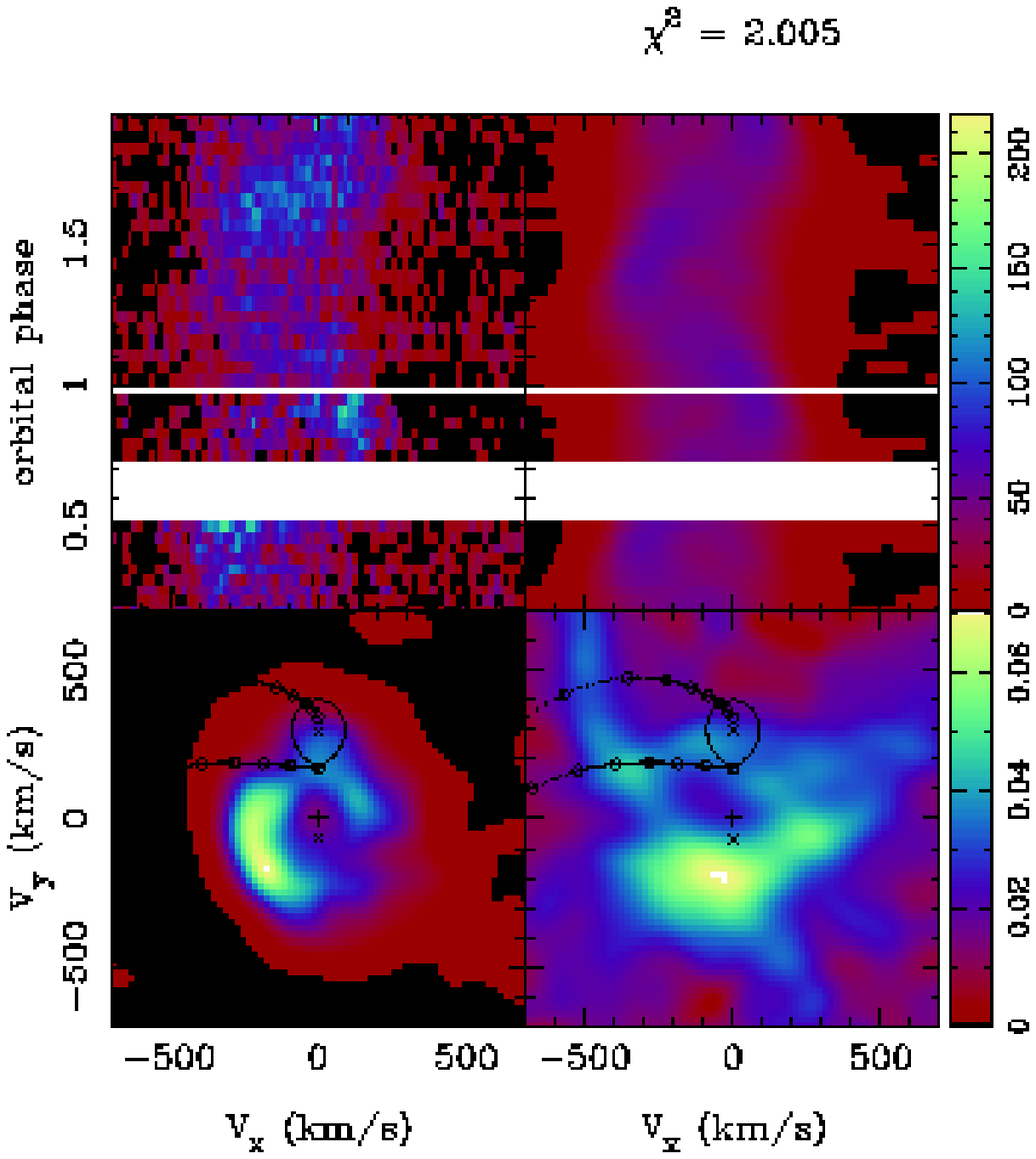}
 \includegraphics[width=80mm]{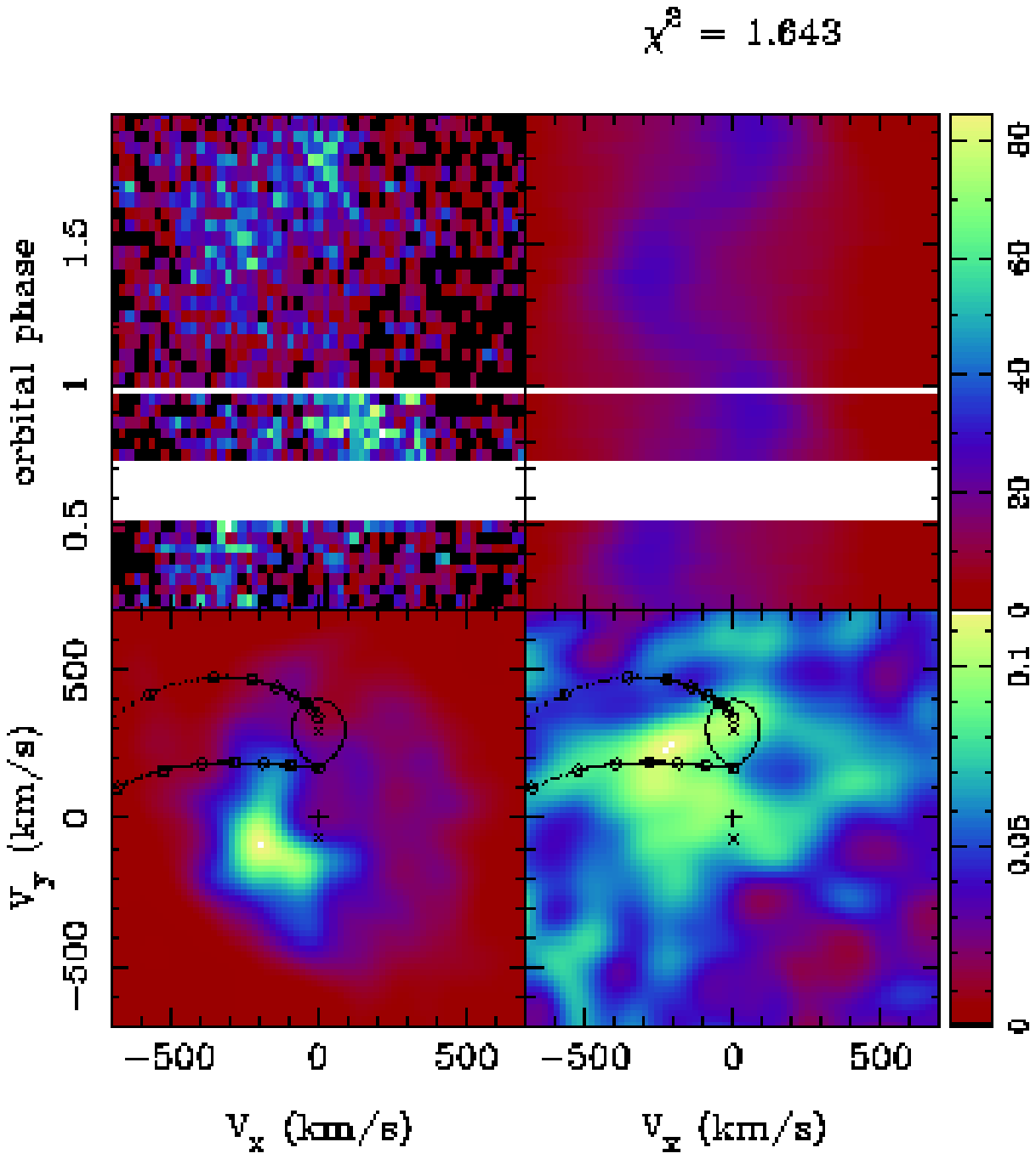}
 \hspace{0.7cm}
 \caption{Modulation tomograms.   In
 each
set of four panels: the observed data (top left) are well reproduced by the fitted
data (top right); the lower left-hand panel shows the constant 
emission of the
disk with a strong hot spot; the lower right-hand panel illustrates the
modulation amplitude of the emission.
Overplotted on the tomograms are the secondary Roche lobe, predicted
primary, secondary, and center-of-mass positions (crosses).  The lower 
curved
line
represents the accretion stream ballistic trajectories
and the upper curved line represents the Keplerian velocity of the disk
along the stream.  The crosses along the trajectories represent
steps of 0.1 R$_{L1}$ (where R$_{L1}$ is the distance from the compact
object to the inner Lagrangian point)
from the primary and asterisks show the apsides of
the accretion stream.
{\bf Left (a):} H$\alpha$ using 15 spectra from June 5, 2010 and
26 from June 26, 2011. {\bf Right (b):}
He~II $\lambda$4686 using 15 spectra from June 5, 2010 and
26 from June 26, 2011.}
\label{mod1}
\end{figure*}

\begin{figure*}
\includegraphics[width=88mm]{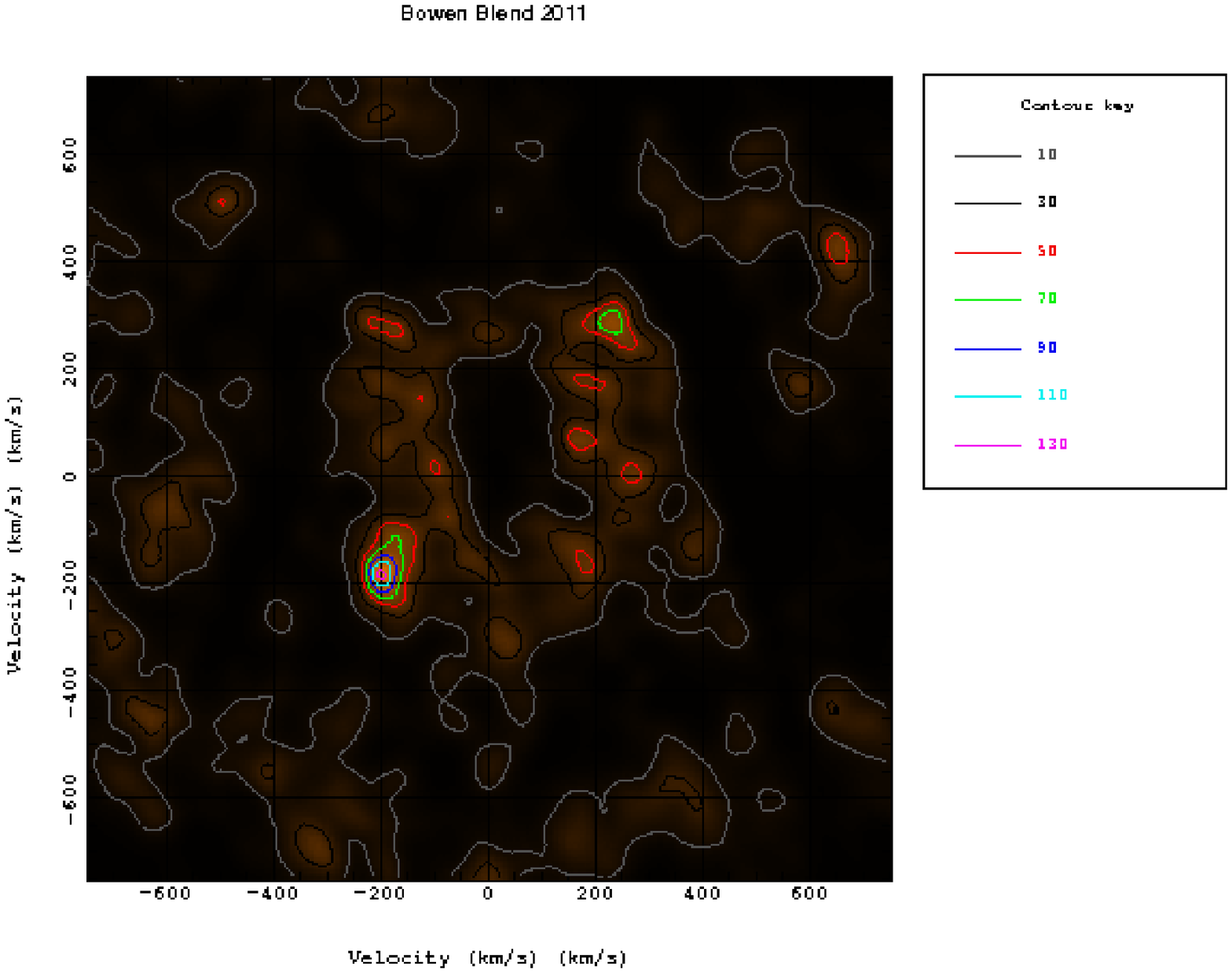}
\includegraphics[width=72mm]{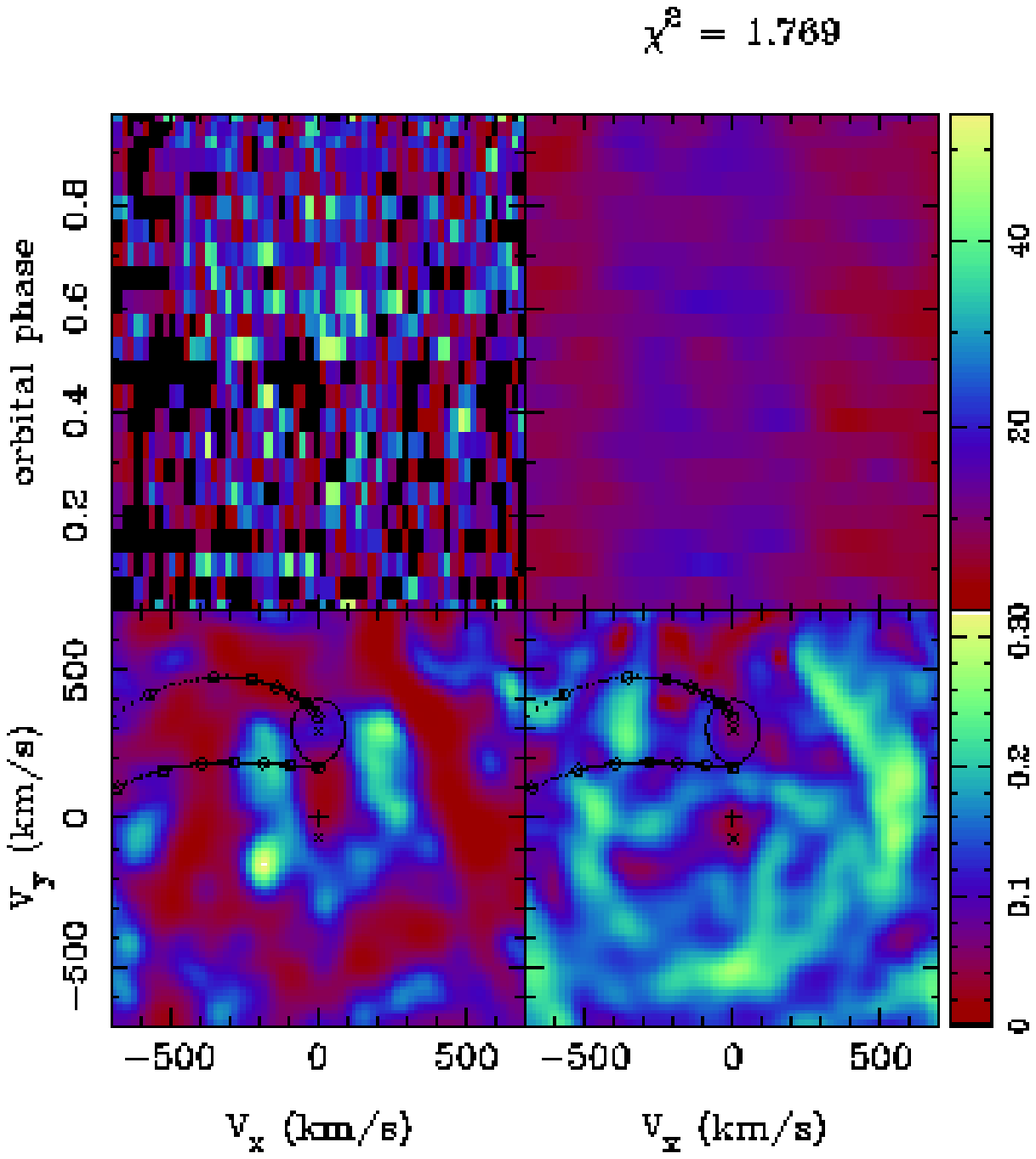}
\caption{Tomograms of the N~III~$\lambda$4640
component of the Bowen complex from 2011. 
{\bf Left (a):} Doppler tomogram. 
{\bf Right (b):} Modulation tomogram.  
}
\label{bowdop}
\end{figure*}

\begin{figure}
\includegraphics[width=80mm,angle=0]{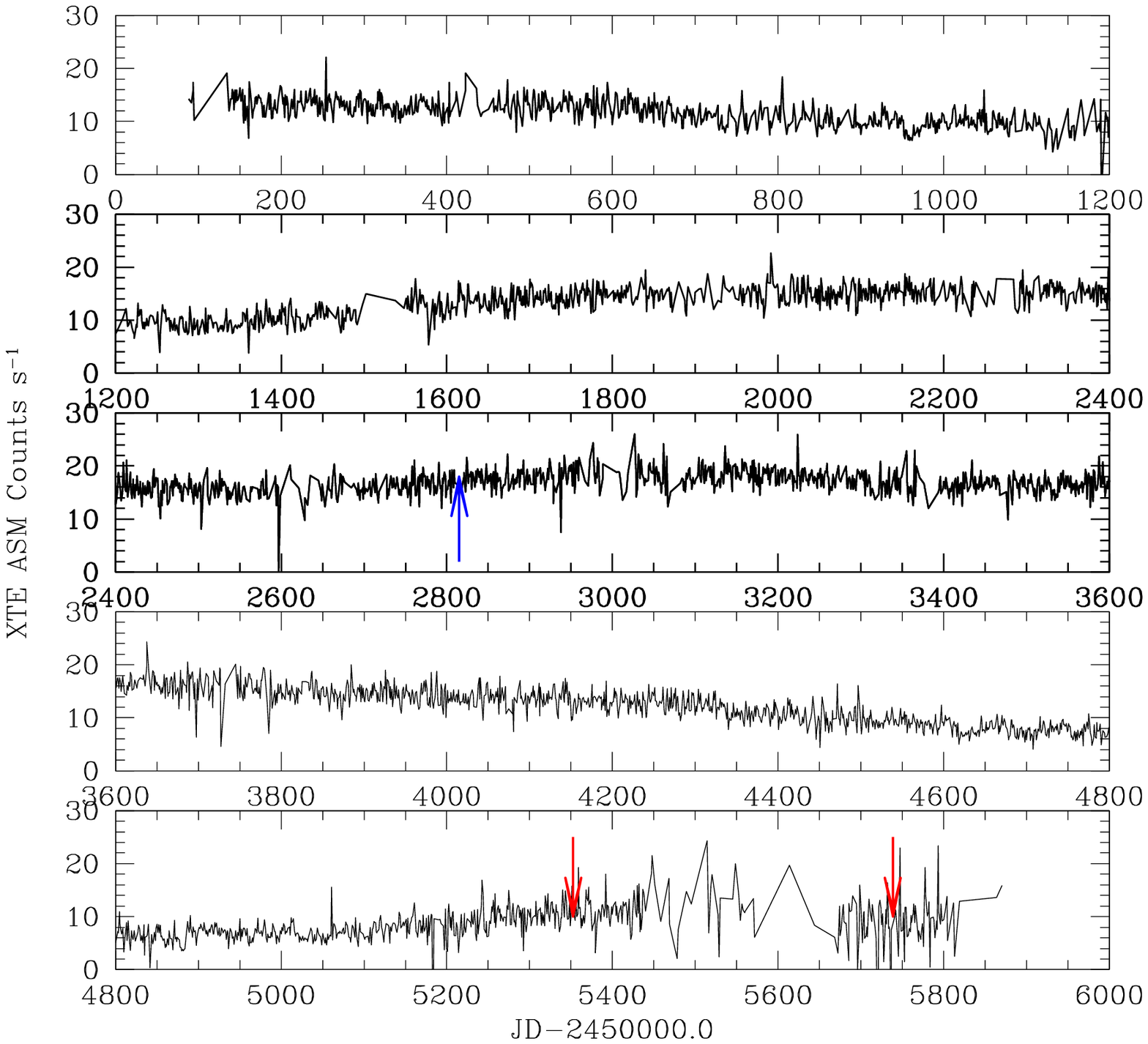}
 \caption{
The {\it RXTE/ASM} light curve of V926 Sco.  Marked in blue is the time of optical observations taken in June 2003 by
Casares~\etal~(2006), times of our optical observations, made in June 2010 \& 2011 are marked in red.
The X-ray count rate during the current
observations, 11$\pm$3 cts/sec (2011) and 8$\pm$3 cts/sec (2010) represent upper limits since the gain of the RXTE/ASM was higher
during its last two years of operation (Remillard, personal communication) and are
significantly lower than during the 2003 observation
(18$\pm$2 cts/sec).}
\hspace{0.3cm}
\label{rxte}
\end{figure}

The modulation tomograms we produced are shown in Figs. \ref{mod1}a,b, 
and \ref{bowdop}b.
In each case, the figure is split into four panels. The upper two 
panels contain the trail of the original spectra around the chosen 
emission line on the left and the trail predicted from the final velocity map, 
for which $\chi^2$ is minimized, on the right. Lighter colors indicate higher 
intensities in each case.
Each figure also contains two two-dimensional velocity maps. The lower
left panel is the same as those produced by Doppler tomography,
showing the brightness of the components of the system averaged over the orbital period. The lower right panel shows the amplitude of modulations in the brightness of the components of the system which are harmonic with the orbital period. In this case, bright areas indicate the regions with the greatest amplitude of modulation.

Each of the tomograms is overlaid with estimated values for the velocities of the neutron star, the donor's Roche Lobe and the center of mass of the system.
We used $K_2$ = 298$\pm$83 km s$^{-1}$  as determined by Casares~\etal (2006). 
Since Casares~\etal~ suggested a range (0.05-0.41)
of mass ratios for V926 Sco we adopted their
middle value of 0.23, giving a $K_1$ of 68 km s$^{-1}$.
We tried both our derived value of $-109\pm 4 km s^{-1}$ and the value
used by Casares~\etal~(2006) 
($-140\pm4 km s^{-1}$) for the systemic velocity. Tests
using a range of values near these velocities favored our
derived value for the best fits.
The Keplerian velocity of the disk along the accretion stream and the ballistic trajectory of the stream are also plotted, with circles along each line indicating steps of 0.1 Lagrangian radii from the compact object.

Because the time-averaged H$\alpha$
and He~II $\lambda$4686 showed little change in FWHM and EW between
June 5, 2010 and June 26, 2011, we were able to combine the
data to improve our modulation tomograms.
Doppler tomograms for both years are consistent with the
modulation tomograms but show less detail, so we present only the
modulation tomograms in this paper.
Both lines show significant enhancement of emission
in the lower left quadrant of the lower left panel (Figs. \ref{mod1}a,b);
the corresponding modulation maps also
show strong variation around these regions as would be expected of
any non-symmetric feature in the system.
The enhanced regions are superposed on crescent shaped
emission with the H$\alpha$ tomogram showing a second crescent
in the upper right quadrant of the lower left panel.

Since the EW of the Bowen complex changed by a factor of two from 2010 to
2011 we made separate tomograms for the two years.
Because of the complexity and weakness of the Bowen blend, we
first constructed Doppler tomograms.
When the complex was stronger (2010) we had 0.9 phase coverage in
only 12 bins, which led to poor maps and indeed showed only
noise.  In 2011, although the complex was weak, we had full phase
coverage with over 20 phase bins (Fig. \ref{bowdop}a), and found a hint
of emission consistent with the disk bulge but no emission
associated with the secondary as reported by Casares~\etal~(2006). 
A modulation tomogram of the 2011 data shows
the same behavior as the Doppler map for 2011 (Fig. \ref{bowdop}b); 
however, as the phased spectra in Fig. \ref{phasecur}c demonstrate, there is no
variation in amplitude within errors of measurement.

\section{Discussion and Conclusions}

Modulation tomography of the H$\alpha$ and He~II $\lambda$4686 
emission lines in the spectra of V926 Sco were found to support the suggestion of earlier
authors (Casares~\etal~2006; Augusteijn~\etal~1998; Smale
\& Corbet 1991) that the accretion disk around the primary contains a large, extended bright region, attributed to a bulge in the disk.

Our observations show significant changes in the disk and secondary
star emission from V926 Sco, both since the 2003 observations of
Casares~\etal~(2006) and between 2010 and 2011 in our own observations.
We find that the FWHM of the Bowen complex, He~II $\lambda$4686, and
H$\beta$
are lower by a factor of about two compared to those reported by
Casares~\etal~(2006).  This suggests significantly lower velocities
present in the system.
While we only have
EW in instrumental counts, we find that the EW of the Bowen complex
is of the same order or considerably less than the EW of 
He~II $\lambda$4686, 
whereas Casares~\etal~found the EW of the Bowen complex to be twice
that of He~II in 2003.

A significant reduction in X-ray emission is seen over the same period in the {\it RXTE/ASM} light curve of the system; the X-ray flux has decreased from an average of approximately 18$\pm$2 cts s$^{-1}$ in 2003 to approximately 11$\pm$3 cts s$^{-1}$ in 2010 and 8$\pm$3 cts s$^{-1}$ in 2011 (see Fig. \ref{rxte}). In
particular, since the 2010-2011 data were obtained during the last two years of the operation of {\it RXTE}, during which it is known that the {\it ASM} was running at a higher gain than that used for calibration, the real count rate is likely to be even lower. This indicates a reduction in the accretion rate of the system.
Since Bowen fluorescence of the secondary is attributed to UV heating of the
secondary surface by irradiation, the lower accretion rate is consistent with
there being lower flux from the disk and compact object and hence
less illumination to produce Bowen from the
secondary.

In both H$\alpha$ and He~II $\lambda$4686 we see crescent
shaped emission as seen by Casares~\etal~(2006) from He~II.
This suggests the possibility of an eccentric disk,
a phenomenon
that occurs at mass ratios below $\sim 0.33$ (Haswell~\etal~2001). 
Normally, the comparatively large mass of the compact object, together with conservation 
of angular momentum, leads to the formation of a circular accretion disk centered 
on the compact object. Below this mass ratio, however, the 3:1 resonance has been found 
to cause eccentric instability, leading to a non-axisymmetric, precessing disk 
(Haswell~\etal~2001). In these cases, we would expect so called ‘superhumps’ to be observed in the light curve of the system
(Calvelo~\etal~2009; Zurita~\etal~2002). While
superhumps
have not been detected in photometric studies of V926 Sco (e.g.
Corbet~\etal~1986; Pederson, van Paradijs, \& Lewin~1981)
an eccentric disk cannot be ruled out. Higher resolution observations
at different epochs are required to either confirm or refute this possibility.
The offset crescent shapes seen in the strong H$\alpha$ tomogram
may also suggest that angular momentum in the
disk is being transported by density waves in the disk.
This behavior has been seen in tomography of 
certain Cataclysmic Variables
(Steeghs~\etal~2003) and
references therein).  Steeghs~\etal~(2000) showed that 
two-armed trailing spirals in position coordinates map into 
two-armed crescent shapes in velocity coordinates.




\acknowledgments

We would like to thank the Las Campanas Observatory for the use of 
the Baade Telescope and the RXTE/ASM team for the 15-year X-ray 
lightcurve of
V926 Sco. We would like to thank our referee for helpful comments.  
We gratefully acknowledge the use of the {\it MOLLY} and 
{\it DOPPLER} software written by T.R. Marsh and the {\it MODMAP} 
software written by D. Steeghs. Funded in part by a 
Smithsonian Institution Endowment Grant to SDV.



{\it Facilities:} \facility{The Walter Baade Telescope 
at Las Campanas Observatory}, \facility{
The All Sky Monitor on the Rossi X-ray Timing Explorer (RXTE/ASM)}.



\appendix




\centerline{\bf References}
\vskip 0.2in
Augusteijn, T., van der Hooft, F., de Jong, J.A., van Kerwijk, M.H., van Paradijs, J., 1998, A\&A, 332, 561

Calvelo, D.E., Vrtilek, S.D., Steeghs, D., Torres, M.A.P., Neilsen, J., Filippenko, A.V., Gonzalez Hernandez, J.I., 2009, MNRAS, 399, 539

Casares, J., Cornelisse, R., Steeghs, D., Charles, P.A., Hynes, R.I.,  O'Brien, K., Strohmayar, T.E., 2006, MNRAS, 373, 1235

Corbet, R.H.D., Thorstensen, J.R., Charles, P.A., Menzies, J.W., Naylor, T., Smale, A.P., 1986, MNRAS, 222, 15

Cowley, A.P., Schmidtke, P.C., Hutchings, J.B.,
\& Crampton, D. 2003, ApJ, 125, 2163

Hasinger, G., van der Klis, M., 1989, A\&A, 225, 79

Haswell, C.A., King, A.R., Murray, J.R., Charles, P.A., 2001,
MNRAS, 321, 475

Hellier, C., Mason, K.O., 1989, MNRAS, 239, 715

Horne, K., Marsh, T.R. 1986, MNRAS, 218, 716

Marsh, T.R., Horne, K., 1988, MNRAS 235, 269

Massey, P., Valdes, F., Barnes, J., 1992, 'A User's Guide to Reducing Slit Spectra with IRAF', (http://iraf.noao.edu/iraf/docs/)

McClintock, J.E., Canizares, C.R., Tarter, C.B., 1975, ApJ, 198, 641

Narayan, R., Nitayananda, R., 1986, ARA\&A, 24, 127

van Paradijs, J., van der Klis, M., Pederson, H., 1988, A\&A, 76, 185

Pederson, H., van Paradijs, J.,Lewin, W.H.G., 1981, Nature, 294, 725

Smale, A.P., \& Corbet, R.H.D., 1991, ApJ, 383, 853

Smale, A.P., Charles, P.A., Tuohy, I.R., Thorstensen, J.R., 1984, MNRAS, 207, 29

Steeghs, D., 2003, MNRAS, 344, 448

Steeghs, D., Horne, K., Harlaftis, E. T., Stehle, R., 2000,
NewAR, 44, 13

Zurita, C., Casares, J., Shahbaz, T., Wagner, R.M., Foltz, C.B., Rodrguez-Gil, P., Hynes, R.I., Charles, P.A., Ryan, E., Schwarz, G., Starrfield, S.G., 2002, MNRAS, 333, 791

\end{document}